%% file: main_final.tex
\documentclass{article}

\usepackage[a4paper, left=3cm, right=3cm, top=3cm]{geometry}
\usepackage{amsmath, amsfonts, amssymb, mathtools, amsthm}
\usepackage{bbm}
\usepackage{hyperref}
\usepackage{graphicx}
\usepackage{ulem}
\usepackage{xcolor}
\usepackage{soul}
\usepackage{cite}
\usepackage{float}
\usepackage{caption}
\usepackage{subcaption} 
\usepackage{layouts}
\usepackage{comment}
\usepackage{authblk}
\usepackage{multirow}

\newcommand{\wsd}{w_{\text{SD}}}
\newcommand{\hatwsd}{\widehat{w}_{\text{SD}}}

\title{A statistical framework for planning and analysing test-retest studies for repeatability of quantitative biomarker measurements}

\author[1,*]{Moritz Fabian Danzer}
\author[1]{Maria Eveslage}
\author[1]{Dennis Görlich}
\author[1, 2, 3, 4]{Benjamin Noto}

\affil[1]{Institute of Biostatistics and Clinical Research, University of Münster, Münster, 48149, Germany}
\affil[2]{Clinic for Radiology, University Hospital Münster, Münster, 48149, Germany}
\affil[3]{Department of Nuclear Medicine, University Hospital Münster, Münster, 48149, Germany}
\affil[4]{West German Cancer Centre (WTZ) Essen-Münster – Münster site, University Hospital Münster, Münster, 48149, Germany}

\affil[*]{moritzfabian.danzer@ukmuenster.de}

\date{}

\begin{document}

\flushbottom
\maketitle
%
%
\thispagestyle{empty}

\begin{abstract}
	\input{abstract.tex}
\end{abstract}

\section{Introduction}
\input{introduction}
\section{Definitions}
\input{definitions}
\section{Effective specificity as a criterion for sample size estimation}\label{sec:effective_specificity}
\input{effective_specificity.tex}
\section{Consideration of effective sensitivity}\label{sec:effective_sensitivit}
\input{sensitivity.tex}
\section{Application example}\label{sec:application_example} 
\input{application}

\section{Discussion}
\input{discussion.tex}


\section*{Acknowledgements}

B.N. was funded as a clinician scientist by the Medical Faculty, University of Münster, Germany. There was no dedicated funding for this study.

\section*{Additional information}
\textbf{Data availability}
No datasets were generated or analysed during the current study. Implementations of exact and approximate formulas can be found in the Supplementary R Code. Additionally, we provide sample sizes based on formula \eqref{eq:n_effective_specificity_exact} in Supplementary Tables S1-S4 generated using our R code.\\
\textbf{Competing interests} 
The authors declare that they have no conflict of interest.

\bibliography{references}
\bibliographystyle{plain}

\section*{Supplementary Material}

\setcounter{table}{0}
\renewcommand{\tablename}{Supplementary Table} 
\renewcommand{\thetable}{S\arabic{table}}

In the following tables, we want to give sample sizes based on formula \eqref{eq:n_effective_specificity_exact} in our main manuscript for different choices of parameters $p_{\text{conf}}$, $p_{\text{esp,lb}}$, $p_{\text{sp}}$ and $m$. For further constellations that cannot be found in the following tables, we would like to refer to the sample size function provided in our Supplementary R Code.

\begin{table}[h]
	\centering
	\caption{Sample sizes for different constellations of $p_{\text{conf}}$, $p_{\text{esp,lb}}$ and $p_{\text{sp}}$ for $m=2$}
	\begin{tabular}{|c|c|c|c|c|c|c|c|}
		\hline
		\multicolumn{8}{|c|}{m=2}\\
		\hline
		\hline
		\multirow{2}{*}{$p_{\text{conf}}$} & \multirow{2}{*}{$p_{\text{esp,lb}}$} & \multicolumn{6}{|c|}{$p_{\text{sp}}$}\\
		\cline{3-8}
		&& 0.800 & 0.900 & 0.925 & 0.950 & 0.975 & 0.990\\
		\hline
		\multirow{6}{*}{ 0.800 } & 0.700 &13 &4 &4 &3 &2 &2 \\
		& 0.800 & &10 &7 &5 &3 &3 \\
		& 0.900 & & &68 &17 &7 &4 \\
		& 0.925 & & & &48 &11 &6 \\
		& 0.950 & & & & &27 &9 \\
		& 0.975 & & & & & &25 \\
		\hline 
		\multirow{6}{*}{ 0.900 } & 0.700 &25 &7 &6 &5 &4 &3 \\
		& 0.800 & &19 &12 &8 &6 &4 \\
		& 0.900 & & &147 &34 &13 &8 \\
		& 0.925 & & & &102 &22 &10 \\
		& 0.950 & & & & &55 &16 \\
		& 0.975 & & & & & &52 \\
		\hline 
		\multirow{6}{*}{ 0.925 } & 0.700 &30 &9 &7 &6 &4 &4 \\
		& 0.800 & &23 &15 &10 &7 &5 \\
		& 0.900 & & &183 &42 &16 &9 \\
		& 0.925 & & & &127 &26 &12 \\
		& 0.950 & & & & &69 &20 \\
		& 0.975 & & & & & &64 \\
		\hline 
		\multirow{6}{*}{ 0.950 } & 0.700 &38 &11 &8 &7 &5 &4 \\
		& 0.800 & &29 &18 &12 &8 &6 \\
		& 0.900 & & &236 &54 &20 &11 \\
		& 0.925 & & & &164 &33 &15 \\
		& 0.950 & & & & &88 &25 \\
		& 0.975 & & & & & &82 \\
		\hline 
		\multirow{6}{*}{ 0.975 } & 0.700 &53 &14 &11 &9 &7 &5 \\
		& 0.800 & &40 &25 &16 &11 &8 \\
		& 0.900 & & &332 &75 &27 &15 \\
		& 0.925 & & & &229 &46 &20 \\
		& 0.950 & & & & &122 &34 \\
		& 0.975 & & & & & &114 \\
		\hline 
		\multirow{6}{*}{ 0.990 } & 0.700 &73 &19 &15 &12 &9 &7 \\
		& 0.800 & &54 &34 &22 &14 &10 \\
		& 0.900 & & &463 &103 &37 &20 \\
		& 0.925 & & & &320 &63 &28 \\
		& 0.950 & & & & &170 &46 \\
		& 0.975 & & & & & &159 \\
		\hline 
	\end{tabular}
\end{table}

\begin{table}
	\centering
	\caption{Sample sizes for different constellations of $p_{\text{conf}}$, $p_{\text{esp,lb}}$ and $p_{\text{sp}}$ for $m=3$}
	\begin{tabular}{|c|c|c|c|c|c|c|c|}
		\hline
		\multicolumn{8}{|c|}{m=3}\\
		\hline
		\hline
		\multirow{2}{*}{$p_{\text{conf}}$} & \multirow{2}{*}{$p_{\text{esp,lb}}$} & \multicolumn{6}{|c|}{$p_{\text{sp}}$}\\
		\cline{3-8}
		&& 0.800 & 0.900 & 0.925 & 0.950 & 0.975 & 0.990\\
		\hline
		\multirow{6}{*}{ 0.800 } & 0.700 &7 &2 &2 &2 &1 &1 \\
		& 0.800 & &5 &4 &3 &2 &2 \\
		& 0.900 & & &34 &9 &4 &2 \\
		& 0.925 & & & &24 &6 &3 \\
		& 0.950 & & & & &14 &5 \\
		& 0.975 & & & & & &13 \\
		\hline 
		\multirow{6}{*}{ 0.900 } & 0.700 &13 &4 &3 &3 &2 &2 \\
		& 0.800 & &10 &6 &4 &3 &2 \\
		& 0.900 & & &74 &17 &7 &4 \\
		& 0.925 & & & &51 &11 &5 \\
		& 0.950 & & & & &28 &8 \\
		& 0.975 & & & & & &26 \\
		\hline 
		\multirow{6}{*}{ 0.925 } & 0.700 &15 &5 &4 &3 &2 &2 \\
		& 0.800 & &12 &8 &5 &4 &3 \\
		& 0.900 & & &92 &21 &8 &5 \\
		& 0.925 & & & &64 &13 &6 \\
		& 0.950 & & & & &35 &10 \\
		& 0.975 & & & & & &32 \\
		\hline 
		\multirow{6}{*}{ 0.950 } & 0.700 &19 &6 &4 &4 &3 &2 \\
		& 0.800 & &15 &9 &6 &4 &3 \\
		& 0.900 & & &118 &27 &10 &6 \\
		& 0.925 & & & &82 &17 &8 \\
		& 0.950 & & & & &44 &13 \\
		& 0.975 & & & & & &41 \\
		\hline 
		\multirow{6}{*}{ 0.975 } & 0.700 &27 &7 &6 &5 &4 &3 \\
		& 0.800 & &20 &13 &8 &6 &4 \\
		& 0.900 & & &166 &38 &14 &8 \\
		& 0.925 & & & &115 &23 &10 \\
		& 0.950 & & & & &61 &17 \\
		& 0.975 & & & & & &57 \\
		\hline 
		\multirow{6}{*}{ 0.990 } & 0.700 &37 &10 &8 &6 &5 &4 \\
		& 0.800 & &27 &17 &11 &7 &5 \\
		& 0.900 & & &232 &52 &19 &10 \\
		& 0.925 & & & &160 &32 &14 \\
		& 0.950 & & & & &85 &23 \\
		& 0.975 & & & & & &80 \\
		\hline 
	\end{tabular}
\end{table}

\begin{table}
	\centering
	\caption{Sample sizes for different constellations of $p_{\text{conf}}$, $p_{\text{esp,lb}}$ and $p_{\text{sp}}$ for $m=4$}
	\begin{tabular}{|c|c|c|c|c|c|c|c|}
		\hline
		\multicolumn{8}{|c|}{m=4}\\
		\hline
		\hline
		\multirow{2}{*}{$p_{\text{conf}}$} & \multirow{2}{*}{$p_{\text{esp,lb}}$} & \multicolumn{6}{|c|}{$p_{\text{sp}}$}\\
		\cline{3-8}
		&& 0.800 & 0.900 & 0.925 & 0.950 & 0.975 & 0.990\\
		\hline
		\multirow{6}{*}{ 0.800 } & 0.700 &5 &2 &2 &1 &1 &1 \\
		& 0.800 & &4 &3 &2 &1 &1 \\
		& 0.900 & & &23 &6 &3 &2 \\
		& 0.925 & & & &16 &4 &2 \\
		& 0.950 & & & & &9 &3 \\
		& 0.975 & & & & & &9 \\
		\hline 
		\multirow{6}{*}{ 0.900 } & 0.700 &9 &3 &2 &2 &2 &1 \\
		& 0.800 & &7 &4 &3 &2 &2 \\
		& 0.900 & & &49 &12 &5 &3 \\
		& 0.925 & & & &34 &8 &4 \\
		& 0.950 & & & & &19 &6 \\
		& 0.975 & & & & & &18 \\
		\hline 
		\multirow{6}{*}{ 0.925 } & 0.700 &10 &3 &3 &2 &2 &2 \\
		& 0.800 & &8 &5 &4 &3 &2 \\
		& 0.900 & & &61 &14 &6 &3 \\
		& 0.925 & & & &43 &9 &4 \\
		& 0.950 & & & & &23 &7 \\
		& 0.975 & & & & & &22 \\
		\hline 
		\multirow{6}{*}{ 0.950 } & 0.700 &13 &4 &3 &3 &2 &2 \\
		& 0.800 & &10 &6 &4 &3 &2 \\
		& 0.900 & & &79 &18 &7 &4 \\
		& 0.925 & & & &55 &11 &5 \\
		& 0.950 & & & & &30 &9 \\
		& 0.975 & & & & & &28 \\
		\hline 
		\multirow{6}{*}{ 0.975 } & 0.700 &18 &5 &4 &3 &3 &2 \\
		& 0.800 & &14 &9 &6 &4 &3 \\
		& 0.900 & & &111 &25 &9 &5 \\
		& 0.925 & & & &77 &16 &7 \\
		& 0.950 & & & & &41 &12 \\
		& 0.975 & & & & & &38 \\
		\hline 
		\multirow{6}{*}{ 0.990 } & 0.700 &25 &7 &5 &4 &3 &3 \\
		& 0.800 & &18 &12 &8 &5 &4 \\
		& 0.900 & & &155 &35 &13 &7 \\
		& 0.925 & & & &107 &21 &10 \\
		& 0.950 & & & & &57 &16 \\
		& 0.975 & & & & & &53 \\
		\hline 
	\end{tabular}
\end{table}

\begin{table}
	\centering
	\caption{Sample sizes for different constellations of $p_{\text{conf}}$, $p_{\text{esp,lb}}$ and $p_{\text{sp}}$ for $m=5$}
	\begin{tabular}{|c|c|c|c|c|c|c|c|}
		\hline
		\multicolumn{8}{|c|}{m=5}\\
		\hline
		\hline
		\multirow{2}{*}{$p_{\text{conf}}$} & \multirow{2}{*}{$p_{\text{esp,lb}}$} & \multicolumn{6}{|c|}{$p_{\text{sp}}$}\\
		\cline{3-8}
		&& 0.800 & 0.900 & 0.925 & 0.950 & 0.975 & 0.990\\
		\hline
		\multirow{6}{*}{ 0.800 } & 0.700 &4 &1 &1 &1 &1 &1 \\
		& 0.800 & &3 &2 &2 &1 &1 \\
		& 0.900 & & &17 &5 &2 &1 \\
		& 0.925 & & & &12 &3 &2 \\
		& 0.950 & & & & &7 &3 \\
		& 0.975 & & & & & &7 \\
		\hline 
		\multirow{6}{*}{ 0.900 } & 0.700 &7 &2 &2 &2 &1 &1 \\
		& 0.800 & &5 &3 &2 &2 &1 \\
		& 0.900 & & &37 &9 &4 &2 \\
		& 0.925 & & & &26 &6 &3 \\
		& 0.950 & & & & &14 &4 \\
		& 0.975 & & & & & &13 \\
		\hline 
		\multirow{6}{*}{ 0.925 } & 0.700 &8 &3 &2 &2 &1 &1 \\
		& 0.800 & &6 &4 &3 &2 &2 \\
		& 0.900 & & &46 &11 &4 &3 \\
		& 0.925 & & & &32 &7 &3 \\
		& 0.950 & & & & &18 &5 \\
		& 0.975 & & & & & &16 \\
		\hline 
		\multirow{6}{*}{ 0.950 } & 0.700 &10 &3 &2 &2 &2 &1 \\
		& 0.800 & &8 &5 &3 &2 &2 \\
		& 0.900 & & &59 &14 &5 &3 \\
		& 0.925 & & & &41 &9 &4 \\
		& 0.950 & & & & &22 &7 \\
		& 0.975 & & & & & &21 \\
		\hline 
		\multirow{6}{*}{ 0.975 } & 0.700 &14 &4 &3 &3 &2 &2 \\
		& 0.800 & &10 &7 &4 &3 &2 \\
		& 0.900 & & &83 &19 &7 &4 \\
		& 0.925 & & & &58 &12 &5 \\
		& 0.950 & & & & &31 &9 \\
		& 0.975 & & & & & &29 \\
		\hline 
		\multirow{6}{*}{ 0.990 } & 0.700 &19 &5 &4 &3 &3 &2 \\
		& 0.800 & &14 &9 &6 &4 &3 \\
		& 0.900 & & &116 &26 &10 &5 \\
		& 0.925 & & & &80 &16 &7 \\
		& 0.950 & & & & &43 &12 \\
		& 0.975 & & & & & &40 \\
		\hline 
	\end{tabular}
\end{table}

\end{document}

%% file: abstract.tex
There is an increasing number of potential biomarkers that could allow for early assessment of treatment response or disease progression. However, measurements of quantitative biomarkers are subject to random variability. Hence, differences of a biomarker in longitudinal measurements do not necessarily represent real change but might be caused by this random measurement variability. Before utilizing a quantitative biomarker in longitudinal studies, it is therefore essential to assess the measurement repeatability. Measurement repeatability obtained from test-retest studies can be quantified by the repeatability coefficient ($RC$), which is then used in the subsequent longitudinal study to determine if a measured difference represents real change or is within the range of expected random measurement variability. The quality of the point estimate of $RC$ therefore directly governs the assessment quality of the longitudinal study.\\
$RC$ estimation accuracy depends on the case number in the test-retest study, but despite its pivotal role, no comprehensive framework for sample size calculation of test-retest studies exists. To address this issue, we have established such a framework, which allows for flexible sample size calculation of test-retest studies, based upon newly introduced criteria concerning assessment quality in the longitudinal study. This also permits retrospective assessment of prior test-retest studies.

%% file: introduction.tex
 	

A biomarker is a characteristic objectively measured and evaluated as an indicator of normal biological processes, pathogenic processes, or response to a therapeutic intervention \cite{biomarkers2001biomarkers}. Biomarkers used as indicators of response to a therapeutic intervention, or disease progression, are called \textit{treatment response biomarkers}.  
One prime, established treatment response biomarker is lesion size change in cross-sectional imaging. For clinical trials concerning solid tumors, the measurement of lesion size is formalized in the so-called Response Evaluation Criteria in Solid Tumors (RECIST) \cite{Eisenhauer2009-je}, that categorize treatment response. 
With the rapid advancement in medical sciences, there is an increasing number of new potential treatment response biomarkers that could possibly allow for early and objective assessment of treatment response or disease progression in clinical trials and clinical practice \cite{Ko2021-sl}.\\
However, using a biomarker in practice requires some basic research into the reliability of its measurement. In addition to a fixed systematic measurement error (bias), which can be investigated by comparing measurements with a known target value (e.g. phantom studies), it is important to take into account that measurements of quantitative biomarkers are subject to random variability.  Hence, changes in a biomarker in longitudinal measurements made under the same conditions do not necessarily represent real change but might be caused by exactly this random measurement variability. Before testing or even utilizing a quantitative biomarker in longitudinal studies, it is therefore of principal importance to assess the measurement repeatability \cite{shukla2019quantitative}.

The repeatability of measurement is determined by test-retest studies, which then are also referred to as repeatability studies. In such studies, replicate measurements are made on a sample of subjects under conditions that are as constant as possible \cite{barnhart2007overview}. Measurement repeatability can be quantified by the within-subject standard deviation ($w_{\text{SD}}$). Using $w_{\text{SD}}$, the repeatability coefficient ($RC$) can be calculated \cite{Obuchowski2018-bo,shukla2019quantitative, bland1996education}.
$RC$ is then used in the longitudinal study to determine if a difference in the biomarker represents presumed real change or is within the range of random measurement variability.
It is defined in such a way that a desired specificity to detect changes -- usually 95\% -- is targeted.\\ 
The $w_{\text{SD}}$ and the $RC$, as determined by the test-retest study, are point estimates, and hence suffer from random error. As we will show, the targeted specificity is therefore generally not achieved in practice.
Following standard statistical results, the more subjects and the more repeated measurements are included in the test-retest study, the more reliable the estimates of $w_{\text{SD}}$ and $RC$ will be. Accordingly, the probability of a relevant deviation of the actually achieved value from the targeted specificity will decrease. The quality of assessments in the longitudinal study and consequently the validity of its results is directly governed by the precision of the estimates of $w_{\text{SD}}$ and $RC$. \\
Of course, exact knowledge of measurement repeatability is not only crucial for biomarkers. 
For example, excellent measurement repeatability of scales and other laboratory instruments is mandatory. The reliability of a scale can be checked using weights with a known mass and it is possible to perform many repeated measurements. In contrast, many biomarkers are measured {\itshape in-vivo}, rendering attainment of large sample sizes difficult.  
Also, it might be necessary from an ethical point of view to keep sample sizes as low as possible, since the measurement in question might be inconvenient, invasive, or even harmful for the patient or the healthy test person. For example, a biomarker might be derived from computed tomography, which involves ionizing radiation. Yet, if the sample size in the test-retest study is small, there is a high chance of obtaining suboptimal estimates of $RC$ with associated detrimental effects on sensitivity and specificity in the longitudinal study.\\
In what follows, we will focus on such and related issues concerning repeatability. Before doing so, note that, related to but different from repeatability is reproducibility. While repeatability represents the measurement precision under constant conditions, i.e, same measurement procedure, same operators, same measuring system, etc., reproducibility is, in contrast, measurement precision under differing conditions as various operators, measuring systems, etc. \cite{kessler2015emerging}. 
\\
Statistical literature concerning requirements for test-retest studies is scarce. One notable study investigating sample size requirements is by Obuchowski and Bullen \cite{Obuchowski2018-bo}. 
In their work, Obuchowski and Bullen conducted a simulation study to investigate the relation between the sample size in the test-retest study and the specificity achieved in a following longitudinal study. The authors give a blanket recommendation for sample size of test-retest studies based on their results from a fixed set of simulation parameters. 
\\
Our goal is to expand upon the results of Obuchowski and Bullen \cite{Obuchowski2018-bo} in several areas. 
First, we want to introduce new quality criteria for the planning of test-retest studies. 
Furthermore, we will expand the considerations to include sensitivity, which has not been investigated in the literature so far. 
Finally, we aim to provide analytical solutions. In contrast to simulation studies, this allows for flexible calculation of sample size requirements and also the retrospective assessment of test-retest studies, as we will show. In doing so, we establish a comprehensive framework in which the notions introduced above are precisely defined.\\ 
In what follows, we will introduce the model used for our framework and study the aspects of specificity and sensitivity in separate sections. Afterwards, we demonstrate the application of our concepts in a practical example and discuss our results. 

%% file: definitions.tex
One possible approach to distinguish true change from random variation in the longitudinal study is to estimate measurement variability in a test-retest study. 
For this purpose, $n$ patients are measured $m$ times within a short period of time, in which their true value presumably does not change. For our considerations we assume independent subjects, e.g. measurement of one target per patient. In addition, independent replicate measurements are necessary, i.e. measurements on a subject need to be made independent of the knowledge of its previous value(s) \cite{bland1999measuring}. Consequently, we establish the following model for the $j$-th measurement of the $i$-th patient $Y_{ij}$ of the test-retest study:
\begin{equation}\label{eq:standard_model}
Y_{ij}=\mu_i + \varepsilon_{ij}
\end{equation}
where $\mu_i$ is the true value for the $i$-th patient and $\varepsilon_{ij}$ is the random error. We assume the random errors to be independent and normally distributed with mean $0$ and variance $w^2_{\text{\text{SD}}}$ \cite{Obuchowski2018-bo}. In particular, it follows that $Y_{ij} \sim \mathcal{N}(\mu_i, \wsd^2)$ for any $i \in \{1,\dots,n\}$ and $j \in \{1,\dots,m\}$ . This model is appropriate when true replicates are studied and a learning effect can be ruled out. As we are only addressing measurement repeatability, a fixed bias does not need to be considered since it cancels out. We also assume that measurement error is independent from the magnitude of $\mu_i$.  From this data, we can estimate the within-patient standard deviation $w_{\text{\text{SD}}}$ \cite{bland1996education} by 
\begin{equation}\label{eq:wSDhat}
\hatwsd \coloneqq \sqrt{\frac{1}{n} \sum_{i=1}^n \frac{1}{m-1} \sum_{j=1}^m (Y_{ij} - \bar{Y}_{i\cdot})^2 },	
\end{equation} 
where $\bar{Y}_{i\cdot} \coloneqq 1/m \sum_{j=1}^m Y_{ij}$ denotes the mean value of the measurements of patient $i$. Following Cochran's theorem \cite{Cochran:1934}, the distribution of this entity is given by
\begin{equation}\label{eq:exact_distribution}
	n(m-1)\frac{\hatwsd^2}{w_{\text{SD}}^2} \sim \chi^2_{n(m-1)}.
\end{equation} 
According to standard asymptotic theory, the following central limit theorem holds for $\hatwsd$:
\begin{equation}\label{eq:clt}
\frac{\hatwsd - \wsd}{ \frac{\wsd}{\sqrt{2n(m-1)}} } \underset{n,m \to \infty}{\overset{\mathcal{D}}{\to}} Z,
\end{equation}
where $Z$ is a standard normally distributed random variable. If the number of repeated measurements differs between subjects, i.e. the $i$-th subject is measured $m_i$ times, the value $n(m-1)$ needs to be replaced by $\sum_{i=1}^n (m_i - 1)$ in all formulas. For the sake of simplicity, we restrict ourselves to the simple case of an equal number of repetitions $m$ per subject.\\
%
In order to assess changes in the measurements of a single patient in the subsequent longitudinal study, the repeatability coefficient ($RC$) is computed \cite{bland1996education}. 
It indicates the range in which two repeated measurements are expected to fall with a certain probability. 
In what follows, we restrict ourselves to the assessment of changes in both directions.  We want to keep our decision rules flexible, i.e. we establish a target specificity $p_{\text{sp}} \in (0,1)$ which shall be reached for patients with no change in their true biomarker value. Hence $RC$ is a function of $p_{\text{sp}}$ and is given by
\begin{equation}\label{eq:RC}
RC(p_{\text{sp}}) \coloneqq \Phi^{-1}(1 - (1-p_{\text{sp}})/2) \cdot \sqrt{2} \cdot \wsd.
\end{equation} 
In most literature the $RC$ is only considered for a fixed targeted specificity of 95\%, i.e. $RC(0.95)$ \cite{shukla2019quantitative,raunig2015quantitative}. \\
In practice, $\wsd$ is unknown and hence replaced by its consistent estimator $\hatwsd$ to obtain the estimated repeatability coefficient 
\begin{equation}
\hat{RC}(p_{\text{sp}}) \coloneqq \Phi^{-1}(1 - (1-p_{\text{sp}})/2) \cdot \sqrt{2} \cdot \hatwsd.
\end{equation}
This quantity can then be applied as cutpoint in the longitudinal study to determine whether there has been change between two consecutive measurements $Y_{\text{pre}}$ and $Y_{\text{post}}$. Here, we also assume, that the measured values have independent errors, but the true levels $\mu_{\text{pre}}$ and $\mu_{\text{post}}$ might actually be different, i.e. we have $Y_{\text{pre}}=\mu_{\text{pre}} + \varepsilon_{\text{pre}}$ and $Y_{\text{post}}=\mu_{\text{post}} + \varepsilon_{\text{post}}$ with $\varepsilon_{\text{pre}}$ and $\varepsilon_{\text{post}}$ being independent and normally distributed with mean $0$ and variance $\wsd^2$.\\
In case the true values have not changed, i.e. $\mu_{\text{pre}}=\mu_{\text{post}}$, the difference $Y_{\text{post}} - Y_{\text{pre}}$ is normally distributed with mean $0$ and variance $2\wsd^2$. Hence, with a probability of $p_{\text{sp}}$, we have $Y_{\text{post}} - Y_{\text{pre}} \in [-RC(p_{\text{sp}}), RC(p_{\text{sp}})]$.\\
The rule to decide whether there is a change for a patient with the two measured values $Y_{\text{pre}}$ and $Y_{\text{post}}$ should thus be whether their difference lies outside or inside the interval $[-RC(p_{\text{sp}}), RC(p_{\text{sp}})]$. As the bounds are unknown in practice, this decision rule is replaced by the decision rule based on the estimated interval $[-\hat{RC}(p_{\text{sp}}), \hat{RC}(p_{\text{sp}})]$. 
Consequently, the targeted specificity $p_{\text{sp}}$ will never be exactly met. This applies analogously to considerations for the sensitivity of this procedure. 

%% file: effective_specificity.tex
Our goal is to quantify the uncertainty introduced by the replacement of $\wsd$ by its estimator $\hatwsd$. As  mentioned, the targeted specificity ($p_{\text{sp}}$) is not met in practice. To assess this problem, we introduce the effective specificity $P_{\text{esp}}$ which is the specificity actually achieved if a realisation of the estimate $\hatwsd$ is plugged in. Hence, $P_{\text{esp}}$ is a random quantity as it depends on the value of $\hatwsd$. We use a capital letter to emphasise that it is indeed a random variable. It can be implicitly defined via
\begin{equation}\label{eq:implicit_definition}
RC(P_{\text{esp}})=\hat{RC}(p_{\text{sp}}).
\end{equation}
Although this quantity is unknown in practice, we can nevertheless analyse its distribution. Firstly, we can compute the expected value $\mathbb{E}[P_{\text{esp}}]$ and the bias, i.e. the difference $\mathbb{E}[P_{\text{esp}}] - p_{\text{sp}}$. This is also the quantity targeted by Obuchowski and Bullen\cite{Obuchowski2018-bo}. Their quality criterion requires $|\mathbb{E}[P_{\text{esp}}] - p_{\text{sp}}|$ to be smaller than 0.01, i.e. they want the mean effective specificity  to deviate less than 1 percentage point from the target specificity, which they set to 95\%.
But what is even more important, from our point of view, is that we can compute quantiles of the distribution of $P_{\text{esp}}$ which will enable us to establish quality guarantees on the effective specificity of the longitudinal studies based on the design parameters $n$ and $m$ of the test-retest study.

\subsection*{Expected value and bias}
According to \eqref{eq:implicit_definition}, $P_{\text{esp}}$ is given by
\begin{align}
P_{\text{esp}}&=RC^{-1}(\hat{RC}(p_{\text{sp}}))\\
&=1 - 2 \cdot \left(1 - \Phi\left( \Phi^{-1}\left( 1 - \frac{1-p_{\text{sp}}}{2} \right) \cdot \frac{\hatwsd}{\wsd}\right)\right).
\end{align}
The function $RC$ can be inverted as it is a continuous, monotonically increasing function on $(0,1)$. The expectation of this random quantity can be computed exactly using \eqref{eq:exact_distribution} or approximately using the central limit theorem \eqref{eq:clt}, according to which the distribution of $\hatwsd/\wsd$ can be approximated with a normal distribution with expectation $1$ and variance $1/(2n(m-1))$. 
Hence, we get
\begin{align}
&\mathbb{E}[P_{\text{esp}}] \nonumber \\
=&1 - 2 \cdot \left(1 - \int_{0}^{\infty} \Phi\left( \Phi^{-1}\left( 1 - \frac{1-p_{\text{sp}}}{2} \right) \cdot w  \right) f_{\chi^2_{n(m-1)}}(n(m-1)w^2)\, 2wn(m-1)\, dw \right) \label{eq:exact_bias} \\
\approx & 1 - 2 \cdot \left(1 - \int_{-\infty}^{\infty} \Phi\left( \Phi^{-1}\left( 1 - \frac{1-p_{\text{sp}}}{2} \right) \cdot w  \right) \sqrt{\frac{n(m-1)}{\pi}} \exp\left(-n(m-1)(w-1)^2 \right) dw \right) \label{eq:asymptotical_bias} 
\end{align}
where $f_{\chi^2_{n(m-1)}}$ denotes the probability density function (PDF) of a $\chi^2$-distributed random variable with $n(m-1)$ degrees of freedom. By numerical evaluation of the terms in \eqref{eq:exact_bias} and \eqref{eq:asymptotical_bias}, the bias can be computed. 


\subsection*{Quantiles of the distribution of $P_{\text{esp}}$}
\begin{figure}
 \includegraphics[width=1\textwidth]{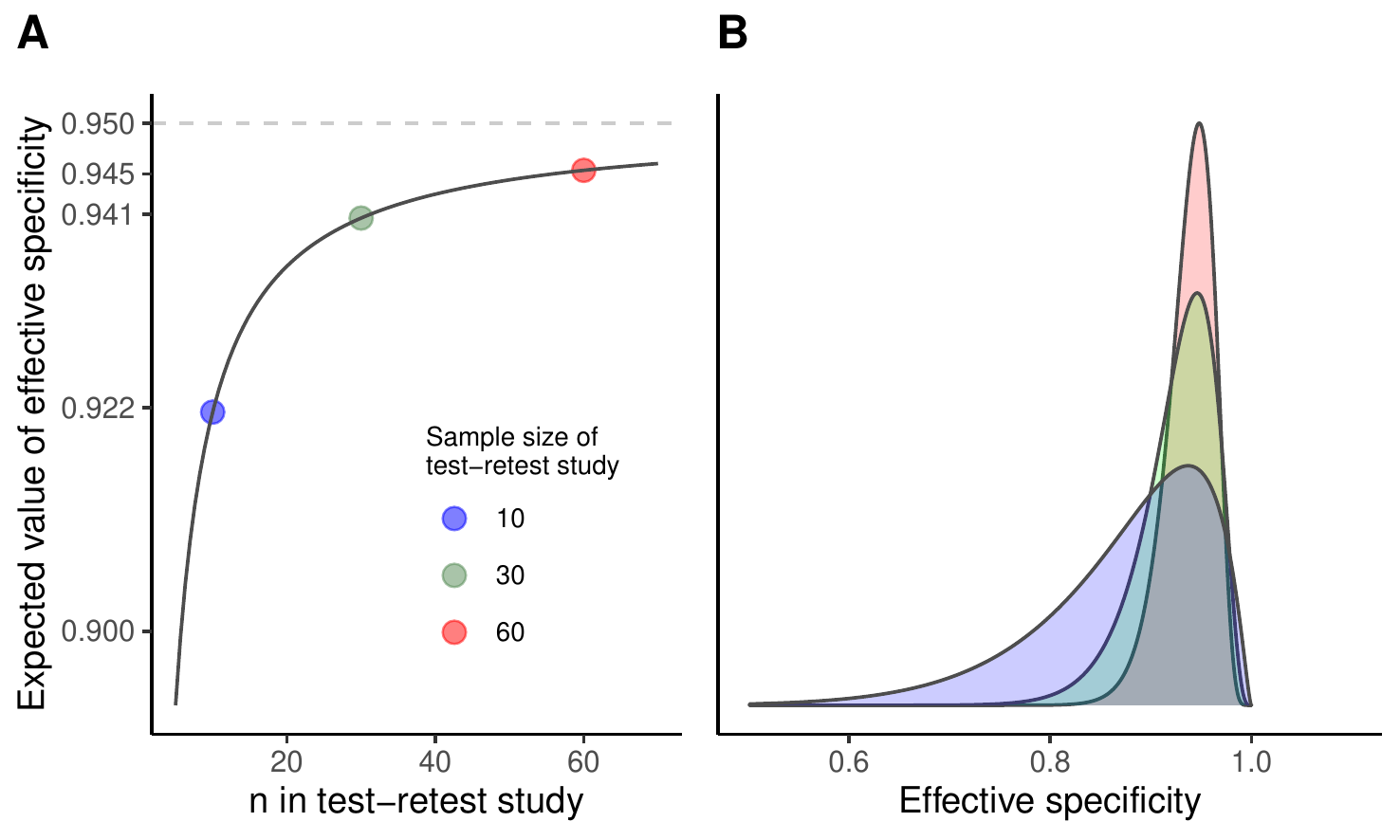}
 \caption{A) Expected value of the effective specificity ($\mathbb{E}[P_{\text{esp}}]$) as a function of  $n$ in the test-retest study for a target specificity $p_{\text{sp}}$ of 95\% and $m=2$. Already for a small case number of 10 (blue dot) the expected value is comparably high. For a case number of 30 (green dot) the bias ($\mathbb{E}[P_{\text{esp}}]-p_{\text{sp}}$) is below 1 percentage point and does not change substantially with an increase of the case number to 60 (red dot).  B) However, while $\mathbb{E}[P_{\text{esp}}]$ is already relatively high for a case number of 10, the tails of the corresponding PDF (blue area) are prominent, resulting in a high chance of obtaining a low $P_{\text{esp}}$ in practice. The green and red area represent the PDF of the effective specificity for $n=30$ and $n=60$, respectively.} 
 \label{fig:bias_and_tails}
\end{figure}

We need to be aware that even if  $\mathbb{E}[P_{\text{esp}}]$ is close to $p_{\text{sp}}$, i.e. the bias is low, the probability for a substantial deviation of the actually realized specificity from the targeted specificity might be large (Figure \ref{fig:bias_and_tails}). 

Therefore, we want to know with which confidence $p_{\text{conf}}$ we can say that the effective specificity is larger than some lower bound $p_{\text{esp,lb}}$. 
This is expressed by the formula
\begin{equation}
\begin{split}
&\mathbb{P}\left[P_{\text{esp}} \geq p_{\text{esp,lb}} \right] = p_{\text{conf}}\\
\Leftrightarrow&\mathbb{P}\left[ \hat{RC}(p_{\text{sp}}) \geq RC(p_{\text{esp,lb}}) \right] = p_{\text{conf}}.
\end{split}
\end{equation}
We want to introduce a new quality criterion based on this concept.\\
The quantity $p_{\text{conf}}$ is a function of $p_{\text{esp,lb}}$ and of course also depends on $p_{\text{sp}},\,n$ and $m$. For notational convenience, however, we omit those arguments.  After some calculations, one obtains
\begin{align}
p_{\text{conf}}=& 1 - F_{\chi^2_{n(m-1)}}\left( \frac{\Phi^{-1}\left(1 - \frac{1-p_{\text{esp,lb}}}{2}\right)^2}{\Phi^{-1}\left(1 - \frac{1-p_{\text{sp}}}{2}\right)^2} n(m-1) \right) \label{eq:exact_confidence} \\
\approx & 1 - \Phi \left( \left( \frac{\Phi^{-1}\left(1 - \frac{1-p_{\text{esp,lb}}}{2}\right)}{\Phi^{-1}\left(1 - \frac{1-p_{\text{sp}}}{2}\right)} - 1 \right) \sqrt{2n(m-1)} \right), \label{eq:asymp_confidence} 
\end{align}
where $F_{\chi^2_{n(m-1)}}$ denotes the cumulative distribution function of a $\chi^2$-distributed random variable with $n(m-1)$ degrees of freedom. 
This formulas can now be used in different ways. In the above form, one can determine the confidence with which the effective specificity exceeds a fixed bound $p_{\text{esp,lb}}$ with given design parameters $n$ and $m$ of the test-retest study. Analogous considerations can be made for upper bounds by computing the probability of the complementary event.\\
In the planning stage of the test-retest study it could be beneficial to choose the sample size $n$ in such a way that a desired lower bound $p_{\text{esp,lb}}$ is achieved with a prespecified confidence $p_{\text{conf}}$. To this end, the asymptotic formula \eqref{eq:asymp_confidence} can be solved explicitly for $n$:

\begin{equation}\label{eq:n_effective_specificity}
n \geq \frac{1}{2(m-1)} \left(  \frac{\Phi^{-1}(1 - p_{\text{conf}}) \Phi^{-1}\left(1 - \frac{1-p_{\text{sp}}}{2}\right) }{ \Phi^{-1}\left(1 - \frac{1-p_{\text{esp,lb}}}{2}\right) - \Phi^{-1}\left(1 - \frac{1-p_{\text{sp}}}{2}\right)}\right) ^2.
\end{equation}
The exact formula \eqref{eq:exact_confidence} cannot be explicitly solved for $n$. However one can numerically solve
\begin{equation}\label{eq:n_effective_specificity_exact}
	\min \left\{ n\in \mathbb{N} \colon 1 - F_{\chi^2_{n(m-1)}}\left( \frac{\Phi^{-1}\left(1 - \frac{1-p_{\text{esp,lb}}}{2}\right)^2}{\Phi^{-1}\left(1 - \frac{1-p_{\text{sp}}}{2}\right)^2} n(m-1) \right) \geq p_{\text{conf}} \right\}. 
\end{equation}
In 
our application example we will apply these formulas in the planning stage of a hypothetical test-retest study.\\
If one wants to identify the worst possible cases for given $n$ and $m$, one could compute the lower bound of the effective specificity which is reached with confidence $p_{\text{conf}}$:

\begin{align}
p_{\text{esp,lb}} & =  1-2\left(1 - \Phi\left( \sqrt{ \frac{F_{\chi^2_{n(m-1)}}^{-1}(1 - p_{\text{conf}})}{n(m-1)}} \Phi^{-1}\left(1 - \frac{1-p_{\text{sp}}}{2}\right) \right) \right) \label{eq:esp_lb_exact} \\
& \approx 1-2\left(1 - \Phi\left( \frac{\Phi^{-1}(1 - p_{\text{conf}}) \Phi^{-1}\left(1 - \frac{1-p_{\text{sp}}}{2}\right)}{\sqrt{2n(m-1)}} + \Phi^{-1}\left(1 - \frac{1-p_{\text{sp}}}{2}\right) \right) \right). \label{eq:esp_lb_asymp}
\end{align}
Accordingly, in $(1-p_{\text{conf}})\cdot 100\%$ of all cases, the effective specificity will be even lower than the obtained $p_{\text{esp,lb}}$.\\
From our point of view, the probability of exceeding a lower bound $p_{\text{esp,lb}}$ is a valid criterion for evaluating the quality of assessment in a longitudinal study. Different from the expected value of $P_{\text{esp}}$ which has been previously proposed as a quality criterion \cite{Obuchowski2018-bo}, our criterion considers the tails of the distribution of $P_{\text{esp}}$. This allows to bound the probability of strongly deviating from the desired specificity.  


%% file: sensitivity.tex
\begin{figure}[H]
 \centering
 \includegraphics{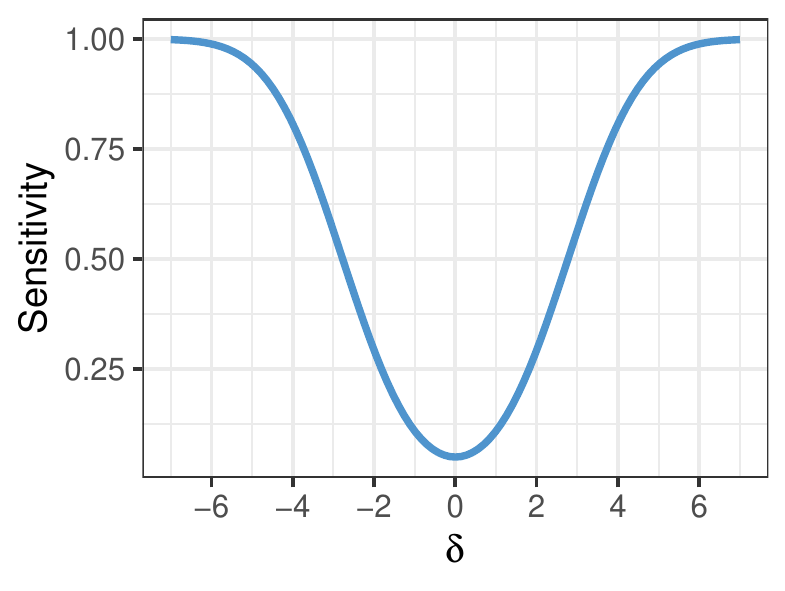}
 \caption{Sensitivity as a function of effect size $\delta \coloneqq \mu_\Delta/\wsd$ for  $p_{\text{sp}}$=0.95. Note that we assume $w_{\text{SD}}$ to be known here.}
 \label{fig:sen_fun_delta}
\end{figure}

Concerning the sensitivity, i.e. the ability to detect real change between two measurements of one patient in the longitudinal study, we can make similar considerations. Before coming back to the problem of the uncertainty caused from the estimation of $\wsd$, we first assume, that $\wsd$ and hence also $RC(p_{\text{sp}})$ is known. Of course, the sensitivity strongly depends on the difference between $\mu_{\text{pre}}$ and $\mu_{\text{post}}$. Also, such differences are more difficult to detect if $\wsd$ is large and a large target specificity is chosen. To be more precise, the sensitivity $p_{\text{se}}$ to detect a difference can be written as a function of $\mu_\Delta\coloneqq \mu_{\text{post}} - \mu_{\text{pre}}$, $\wsd$ and the chosen specificity $p_{\text{sp}}$. It is given by
\begin{equation}
\begin{split}
&p_{\text{se}}(\mu_\Delta, \wsd)\\
\coloneqq&\mathbb{P}[Y_{\text{post}} - Y_{\text{pre}} \notin [-RC(p_{\text{sp}}), RC(p_{\text{sp}})]]\\
=&1 - \left( \Phi\left( \Phi^{-1} \left( 1 - \frac{1 - p_{\text{sp}}}{2} \right) - \frac{\mu_\Delta}{\sqrt{2}\wsd} \right) - \Phi\left(\Phi^{-1} \left(\frac{1 - p_{\text{sp}}}{2} \right) - \frac{\mu_\Delta}{\sqrt{2}\wsd} \right) \right).
\end{split}
\end{equation}
In this form, the function can also be seen as a function of the effect size $\delta \coloneqq \mu_\Delta/\wsd$, i.e.
\begin{equation}\label{eq:sensitivity_delta}
\begin{split}
&p_{\text{se}}(\delta)\\
\coloneqq&1 - \left( \Phi\left( \Phi^{-1} \left( 1 - \frac{1 - p_{\text{sp}}}{2} \right) - \frac{\delta}{\sqrt{2}} \right) - \Phi\left( \Phi^{-1} \left(\frac{1 - p_{\text{sp}}}{2} \right) - \frac{\delta}{\sqrt{2}} \right) \right).
\end{split}
\end{equation}
This dependence of the sensitivity from the effect size $\delta$ is visualized in Figure \ref{fig:sen_fun_delta}.

As $\wsd$ is unknown and needs to be estimated by $\hatwsd$ which will then be plugged in to compute $\hat{RC}(p_{\text{sp}})$, the sensitivity computed in \eqref{eq:sensitivity_delta} will not be reached. Analogously to our considerations for the specificity, we introduce the effective sensitivity $P_{\text{ese}}$ which is the sensitivity which is actually achieved if a realisation of the estimate $\hatwsd$ is plugged in. Of course, it is also a random variable and does depend again on $\mu_\Delta$, $w_{\text{SD}}$ and $p_{\text{sp}}$. It can be defined by the equation
\begin{equation}\label{eq:effective_sensitivity}
\begin{split}
P_\text{ese}(\mu_\Delta, \wsd)&\coloneqq\mathbb{P}[Y_{\text{post}} - Y_{\text{pre}} \notin [-\hat{RC}(p_{\text{sp}}), \hat{RC}(p_{\text{sp}})]|\hatwsd]\\
&=1 - \left( \Phi\left( \Phi^{-1} \left( 1 - \frac{1 - p_{\text{sp}}}{2} \right) \frac{\hatwsd}{\wsd} - \frac{\mu_\Delta}{\sqrt{2}\wsd} \right) \right.\\
&\qquad \left. - \Phi\left(\Phi^{-1} \left( \frac{1-p_{\text{sp}}}{2} \right) \frac{\hatwsd}{\wsd} - \frac{\mu_\Delta}{\sqrt{2}\wsd} \right) \right).
\end{split}
\end{equation}
With this expression and the exact distribution of $\hatwsd$ given as in \eqref{eq:exact_distribution} resp. the approximation of the distribution of $\frac{\hatwsd}{\wsd}$ by a normal distribution from \eqref{eq:clt} we can now quantify the bias caused by the replacement of $\wsd$ by $\hatwsd$ and compute quantiles of the distribution of $P_{\text{ese}}$ which will enable us to also give quality guarantees on the effective sensitivity. Unlike our considerations for the specificity,
these values will also depend from the actual $w_{\text{SD}}$ and the difference $\mu_{\Delta}$ of the longitudinal study and hence will be regarded as functions of those.

\subsection*{Bias}
To compute the bias in dependence from $p_{\text{sp}}$, $\mu_\Delta$ and $\wsd$, we can take the expectation of the right hand side of \eqref{eq:effective_sensitivity} and use the exact distribution \eqref{eq:exact_distribution} and the central limit theorem \eqref{eq:clt} to obtain the result
\begin{equation}
\begin{split}
&\mathbb{E}[P_{\text{ese}}(p_{\text{sp}}, \mu_\Delta, \wsd)]\\
=&1-\int_{0}^{\infty} \Phi\left( \Phi^{-1}\left( 1 - \frac{1-p_{\text{sp}}}{2} \right) \cdot w - \frac{\mu_\Delta}{\sqrt{2}\wsd}  \right) f_{\chi^2_{n(m-1)}}(n(m-1)w^2)2wn(m-1)\, dw\\
&+\int_{0}^{\infty} \Phi\left(\Phi^{-1}\left( \frac{1-p_{\text{sp}}}{2} \right) \cdot w - \frac{\mu_\Delta}{\sqrt{2}\wsd}  \right) f_{\chi^2_{n(m-1)}}(n(m-1)w^2)2wn(m-1)\, dw\\
\approx&1-\int_{-\infty}^{\infty} \Phi\left( \Phi^{-1}\left( 1 - \frac{1-p_{\text{sp}}}{2} \right) \cdot w - \frac{\mu_\Delta}{\sqrt{2}\wsd}  \right) \sqrt{\frac{n(m-1)}{\pi}} \exp\left(-n(m-1)(w-1)^2 \right) dw\\
&+\int_{-\infty}^{\infty} \Phi\left(\Phi^{-1}\left( \frac{1-p_{\text{sp}}}{2} \right) \cdot w - \frac{\mu_\Delta}{\sqrt{2}\wsd}  \right) \sqrt{\frac{n(m-1)}{\pi}} \exp\left(-n(m-1)(w-1)^2 \right) dw.
\end{split}
\end{equation}
Please note that this can essentially be seen as a function of $\delta$. Following \eqref{eq:effective_sensitivity} the bias of the effective sensitivity can be considered as a function of $\delta$ for any given $p_{\text{sp}}$, i.e. $\mathbb{E}[P_{\text{ese}}(p_{\text{sp}}, \delta)] - p_\text{se}(p_{\text{sp}}, \delta)$. 

\subsection*{Quantiles of the distribution of $P_{\text{ese}}$}
For the most accurate examination of the distribution of $P_{\text{ese}}$ we would need to consider both events 
\begin{align}
	&Y_{\text{post}} - Y_{\text{pre}} > \hat{RC}(p_{\text{sp}}) \text{ and }\label{eq:positive_excess}\\
	&Y_{\text{post}} - Y_{\text{pre}} < -\hat{RC}(p_{\text{sp}}).\label{eq:negative_excess}
\end{align}
However, this leads to expressions that are difficult to handle analytically. Actually, the two probabilities
\begin{align}
	&\mathbb{P}[Y_{\text{post}} - Y_{\text{pre}} > \hat{RC}(p_{\text{sp}})|\hatwsd] \text{ and } \label{eq:prob_positive_excess}\\
	&\mathbb{P}[Y_{\text{post}} - Y_{\text{pre}} < -\hat{RC}(p_{\text{sp}})|\hatwsd] \label{eq:prob_negative_excess}
\end{align}
sum up to the effective sensitivity. However, in the presence of an effect, one of them will be much larger than the other. In the case $\delta > 0$, the probability in \eqref{eq:prob_positive_excess} is larger than that from \eqref{eq:prob_negative_excess} which is bounded from above by 0.025 and quickly converges to 0 as $\delta$ increases. To enable the derivation of analytical formulas, we will therefore restrict ourselves to the consideration of $\delta>0$ and the event \eqref{eq:positive_excess}. It is nevertheless possible to circumvent this simplification by numerical inversion of the relationship given in \eqref{eq:effective_sensitivity}. But here, we will approximate
\begin{align}
	P_{\text{ese}}(p_{\text{sp}}, \mu_\Delta, \wsd)&\approx \mathbb{P}[Y_{\text{post}} - Y_{\text{pre}} > \hat{RC}(p_{\text{sp}})|\hatwsd]\\
	&=1 - \Phi\left( \Phi^{-1} \left( 1 - \frac{1 - p_{\text{sp}}}{2} \right) \frac{\hatwsd}{\wsd} - \frac{\mu_\Delta}{\sqrt{2}\wsd} \right).
\end{align}
In analogy to the previous section we can provide confidence levels $p_{\text{conf}}$ which indicate the probability that the effective sensitivity for some effect $\delta$ exceeds the lower bound $p_{\text{ese,lb}}$:
\begin{align}\label{eq:confidence_sensitivity}
\begin{split}
p_{\text{conf}}&=\mathbb{P}[P_{\text{ese}}(p_{\text{sp}}, \delta) \geq p_{\text{ese,lb}}]\\
&\approx F_{\chi^2_{n(m-1)}} \left( \left( \frac{ \Phi^{-1}(1 - p_{\text{ese,lb}}) + \delta/\sqrt{2} }{\Phi^{-1} \left( 1 - \frac{1 - p_{\text{sp}}}{2} \right) }\right)^2 n(m-1) \right)\\
&\approx \Phi \left( \left( \frac{ \Phi^{-1}(1 - p_{\text{ese,lb}}) + \delta/\sqrt{2} }{\Phi^{-1} \left( 1 - \frac{1 - p_{\text{sp}}}{2} \right) } - 1 \right) \sqrt{2n(m-1)} \right)
\end{split}
\end{align}
Of course, such considerations only make sense if $p_{\text{se}} > p_{\text{ese,lb}}$ for the chosen effect size $\delta$. As above, analogous considerations can be made for upper bounds by computing the probability of the complementary event.\\
While \eqref{eq:confidence_sensitivity} allows to compute the confidence of reaching a certain lower bound of the sensitivity for an effect $\delta$, this formula may also be transformed to be used in the planning stage of the test-retest study. If one wants to achieve a fixed confidence with which the effective sensitivity for an effect size $\delta$ exceeds some lower bound, one can use the exact results from above or the approximations made thereafter to determine the sample size $n$ of the test-retest study in which each patient is measured $m$ times. It shall be chosen such that
\begin{align} 
n & = \min \left\{n\in \mathbb{N} \colon \mathbb{P}\left[1 - \left( \Phi\left( \Phi^{-1} \left( 1 - \frac{1 - p_{\text{sp}}}{2} \right) \frac{\hatwsd}{\wsd} - \frac{\mu_\Delta}{\sqrt{2}\wsd} \right) \right. \right. \right.\\
&\qquad \qquad \qquad \qquad \qquad \left. \left. \left. - \Phi\left(\Phi^{-1} \left( \frac{1-p_{\text{sp}}}{2} \right) \frac{\hatwsd}{\wsd} - \frac{\mu_\Delta}{\sqrt{2}\wsd} \right) \right)\geq p_{\text{ese,lb}}\right] \geq p_{\text{conf}} \right\} \\
& \approx \min \left\{n\in \mathbb{N} \colon F_{\chi^2_{n(m-1)}} \left( \left( \frac{ \Phi^{-1}(1 - p_{\text{ese,lb}}) + \delta/\sqrt{2} }{\Phi^{-1} \left( 1 - \frac{1 - p_{\text{sp}}}{2} \right) }\right)^2 n(m-1) \right) \geq p_{\text{conf}} \right\}\\
& \approx \frac{1}{2(m-1)} \left(  \frac{\Phi^{-1}(p_{\text{conf}}) \Phi^{-1}\left(1 - \frac{1-p_{\text{sp}}}{2}\right) }{ \Phi^{-1}\left(1-p_{\text{ese,lb}}\right) +\delta/\sqrt{2} - \Phi^{-1}\left(1 - \frac{1-p_{\text{sp}}}{2}\right)}\right) ^2. \label{eq:n_effective_sensitivity}
\end{align}
Analogous to the preceding section, we can use these results in the planning stage of a test-retest study, as we will demonstrate in the following application example. 
Even if a study is planned based on considerations of the specificity, the formulas allow to assess the distribution of the effective sensitivity for any given effect size of interest.\\
Calculations for $\delta < 0$ follow analogously to the considerations for $\delta > 0$.

%% file: application.tex
\begin{figure}
	\includegraphics[scale=0.8]{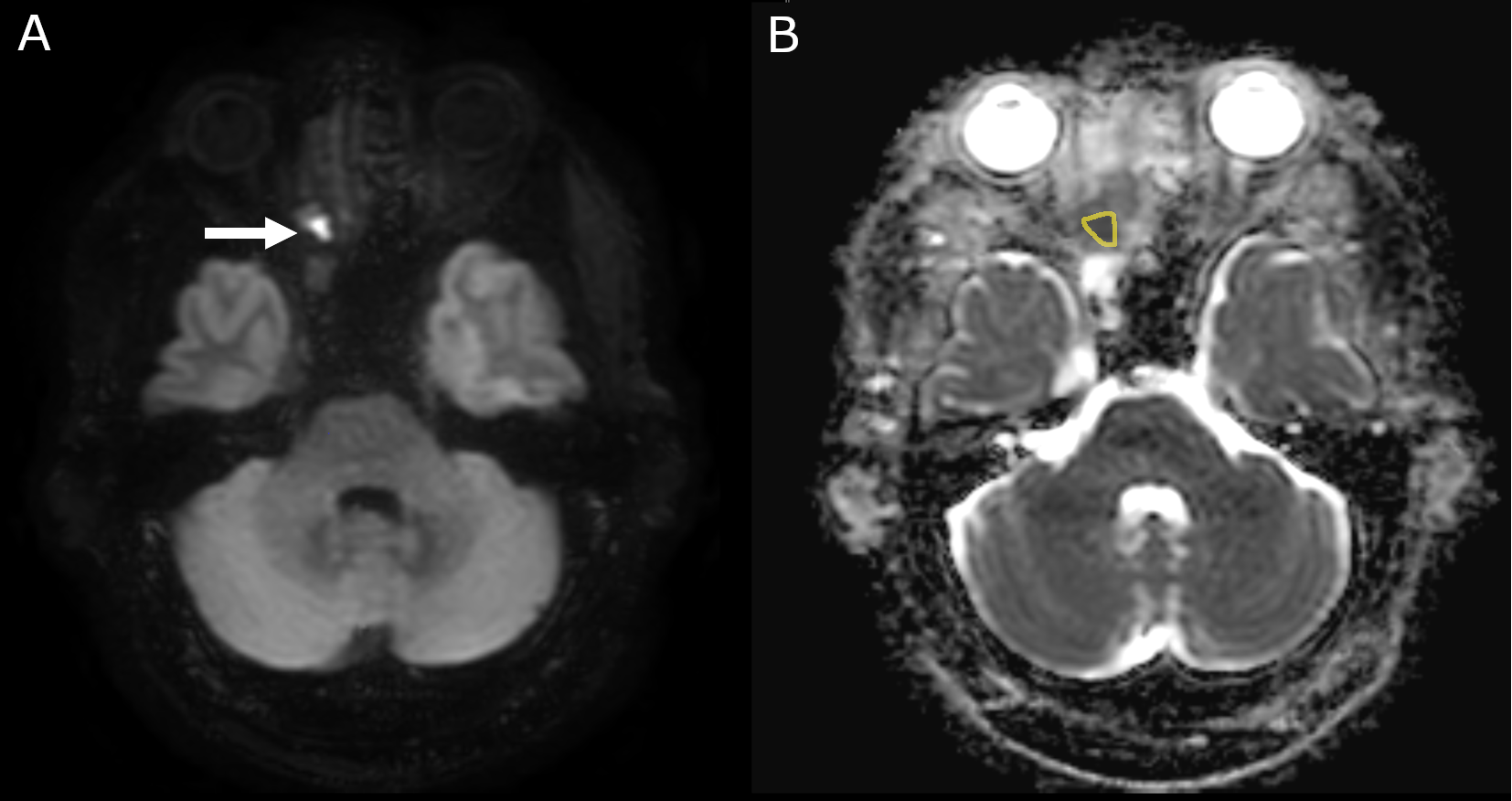}
	\caption{Example case of a tumor in the right nose showing restricted diffusion (A) with a mean ADC of $610 \cdot 10^-6$ $mm^2/s$. (B) Region of interest outlined in yellow.}
	\label{fig:nose}
\end{figure}

To illustrate our considerations, we will discuss a hypothetical application for early treatment response assessment in recurrent or metastatic nasopharyngeal carcinoma. 
While some patients with recurrent nasopharyngeal carcinoma show response or stable disease to systemic treatment, many patients will have progressive disease, which is invariably lethal \cite{glazar2022early, ma2018antitumor}. Nevertheless, futile treatments should be avoided due to associated  toxicity \cite{glazar2022early,chan2005multicenter}. To suspend futile treatment as soon as possible an imaging biomarker is desirable which accurately classifies treatment response earlier than change in morphologic lesion size, the current standard. A promising biomarker in this context is diffusion weighted magnetic resonance imaging (DWI) \cite{lee2021diffusion}. DWI depends on the differences in the movement of water molecules based on Brownian motion, which can be quantified by the apparent diffusion coefficient (ADC). An exemplary measurement of ADC is shown in Figure \ref{fig:nose}. Change in ADC has shown promise as an early treatment response marker in various tumors, including nasopharyngeal carcinoma \cite{lee2021diffusion, torkian2022diffusion, winfield2019utility, winfield2019diffusion}.\\

\subsection*{Prospective planning of test-retest studies}
\begin{figure}
	\centering
	\begin{subfigure}[t]{0.55\textwidth}
		{\Large \textsf{\textbf{A}}}
	\end{subfigure}
	\begin{subfigure}[t]{0.44\textwidth}
		{\Large \textsf{\textbf{B}}}
	\end{subfigure}
	\begin{subfigure}[tb]{0.55\textwidth}
		\includegraphics[width=\linewidth]{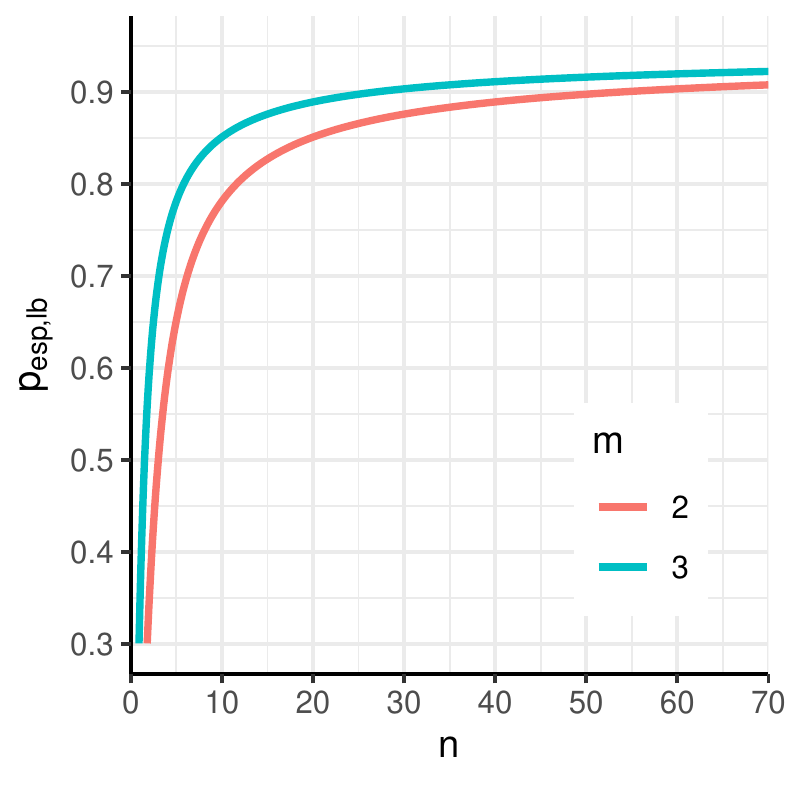}
	\end{subfigure}
	\begin{subfigure}[tb]{0.44\textwidth}
		\centering
		\begin{tabular}{c|c||c|c}
			$p_{\text{esp,lb}}$&m&n&$\mathbb{E}[P_{\text{esp}}]$\\[5pt]
			\hline
			\hline
			0.7&2&7&0.9092\\[5pt]
			0.8&2&12&0.9264\\[5pt]
			0.9&2&54&0.9448\\[5pt]
			0.925&2&164&0.9483\\[5pt]
			\hline
			0.7&3&4&0.9143\\[5pt]
			0.8&3&6&0.9264\\[5pt]
			0.9&3&27&0.9448\\[5pt]
			0.925&3&82&0.9483\\[5pt]
		\end{tabular}
	\end{subfigure}
	\caption{A) Lower bound of effective specificity reached with confidence of 95\% as a function of sample size ($n$) and number of repeated measurements ($m$) of the test-retest study.\\
	B) Sample size resulting from \eqref{eq:n_effective_specificity_exact} for different values of the desired lower bound $p_{\text{esp,lb}}$ that shall be exceeded with a fixed confidence of 95\%.}
	\label{fig:Formel13}
\end{figure}

\begin{figure}[h] 
	\centering
	\includegraphics[width=1\textwidth]{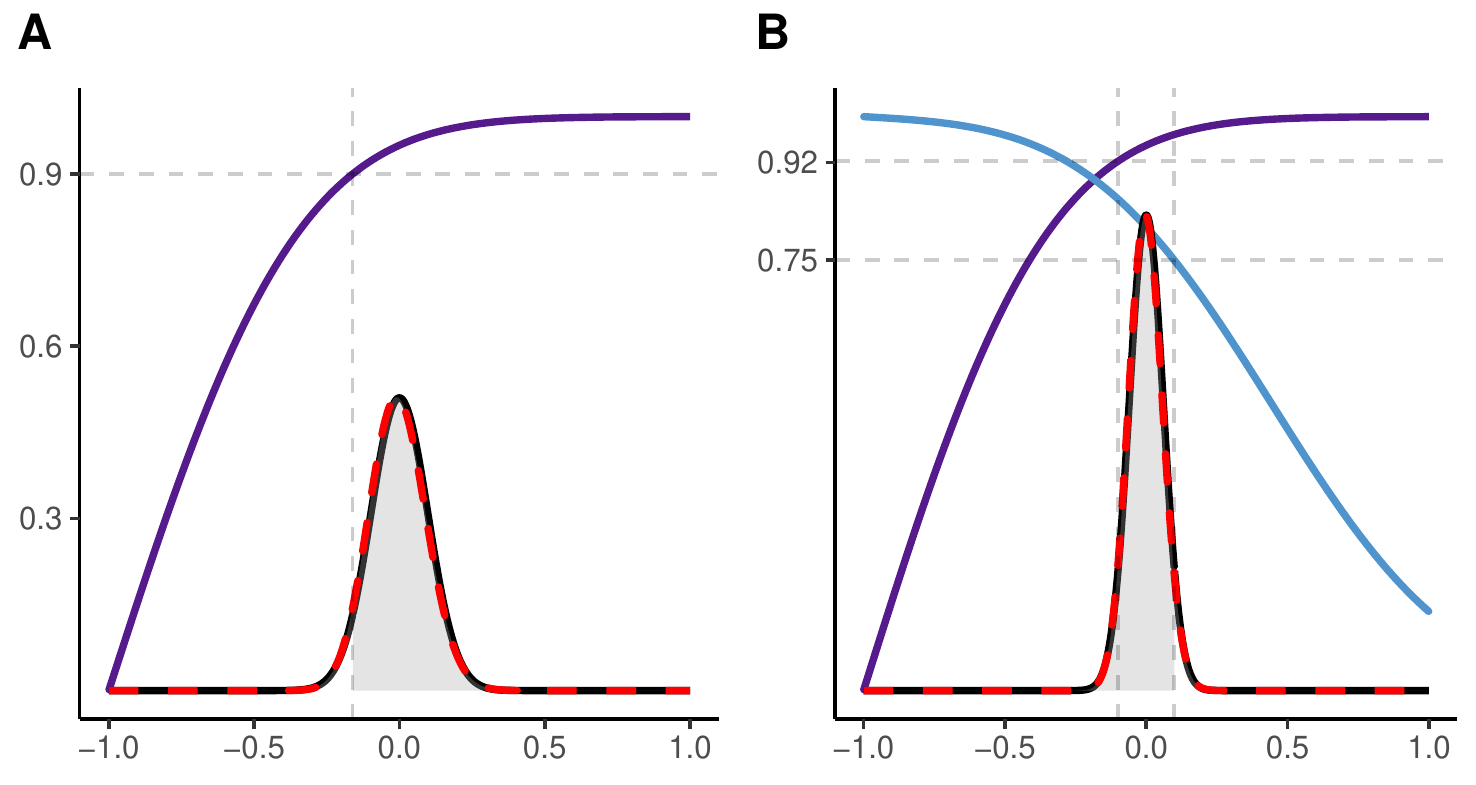}
	\caption{Visualization of the application example. The x-axis denotes the relative error between $\hatwsd$ and $\wsd$. The solid black line represents the asymptotic PDF of the relative error. Note the near identity to the dashed red curve, representing the PDF of the exact $\chi^2$ distribution.
	The violet line shows the effective specificity. Analogously, the blue line shows the effective sensitivity for an underlying effect size of $\delta=4$.\\	
	A) For $n=53$ and $m=2$: The area shaded in light gray represents 95\% of the area under the normal curve. I.e. there is a 95\% chance of obtaining a $\hatwsd$ from the test-retest study that will result in an effective specificity of greater than 90\%.\\
	B) For $n=139$ and $m=2$: The area shaded in light gray represents 95\% of the area under the normal curve. I.e. there is a 95\% chance of obtaining a $\hatwsd$ from the test-retest study that will result in a specificity of greater than 75\%. In this case, an effective specificity of 92.27\% will be reached with a certainty of 95\%.}
	\label{fig:wSDhat_dist}
\end{figure}

As laid out above, before conducting a longitudinal study in which a biomarker is applied to assess treatment response, a test-retest study should be conducted to assess repeatability. 
In our example, we will set $p_\text{sp}$ to 95\% and $m = 2$, as these are the usual values in the literature. We imagine the researcher would want to obtain a specificity of at least 90\% ($p_{\text{esp,lb}}$) with 95\% certainty ($p_{\text{conf}}$) in the longitudinal study. What sample size ($n$) is necessary in the test-retest study? This question can be answered using the asymptotic formula \eqref{eq:n_effective_specificity}:
\begin{equation}
\begin{split}
&n \geq \frac{1}{2(m-1)} \left(  \frac{\Phi^{-1}(1-p_{\text{conf}}) \Phi^{-1}\left(1 - \frac{1-p_{\text{sp}}}{2}\right) }{ \Phi^{-1}\left(1 - \frac{1-p_{\text{esp,lb}}}{2}\right) - \Phi^{-1}\left(1 - \frac{1-p_{\text{sp}}}{2}\right)}\right)^2\\
\Leftrightarrow & n \geq \frac{1}{2(2-1)} \left(  \frac{\Phi^{-1}(0.05) \Phi^{-1}\left(0.975\right) }{ \Phi^{-1}\left(0.95\right) - \Phi^{-1}\left(0.975\right)}\right)^2\\
\Leftrightarrow & n \geq 52.3 
\end{split}
\end{equation}
This can also be concluded from Figure \ref{fig:Formel13} A. Numerical solution of the exact formula \eqref{eq:n_effective_specificity_exact} yields a sample size of 54. Resulting sample sizes for other values of $p_{\text{esp,lb}}$ can be taken from Figure \ref{fig:Formel13} B. The resulting scenario in terms of the distribution of the relative error in the estimation of $\hatwsd$ and its effect on $P_{\text{esp}}$ is displayed in Figure \ref{fig:wSDhat_dist} A.\\

Analogous considerations can be made for the effective sensitivity. We consider the sensitivity for an underlying true effect size of $\delta=4$ in a study with $p_{\text{sp}} = 0.95$ and $m = 2$. According to formula \eqref{eq:sensitivity_delta}, a sensitivity of 80.74\% was achieved if $\wsd$ was a known quantity. However, this will not be met in practice. What is the minimum sample size ($n$) of the test-retest study such that we can be 95\% ($p_{\text{conf}}$) sure to achieve at least a sensitivity of 75\% ($p_{\text{ese,lb}}$) for that effect size? This question can be answered using the approximate formula \eqref{eq:n_effective_sensitivity}.
\begin{align} 
& n \approx \frac{1}{2(m-1)} \left(  \frac{\Phi^{-1}(p_{\text{conf}}) \Phi^{-1}\left(1 - \frac{1-p_{\text{sp}}}{2}\right) }{ \Phi^{-1}\left(1-p_{\text{se,lb}}\right) + \delta/\sqrt{2} - \Phi^{-1}\left(1 - \frac{1-p_{\text{sp}}}{2}\right)}\right)^2 \\
\Rightarrow & n \approx \left(  \frac{\Phi^{-1}(0.95) \Phi^{-1}\left(0.975\right) }{ \Phi^{-1}\left(0.25\right) + 4/\sqrt{2} - \Phi^{-1}\left(0.975\right)}\right)^2\\
\Leftrightarrow & n \approx 138.1
\end{align}
Accordingly, a sample size of 139 patients in the test-retest study would be recommended to achieve the set targets. Using the exact formula \eqref{eq:esp_lb_exact} or the asymptotic formula \eqref{eq:esp_lb_asymp}, an effective specificity of at least 92.25\% resp. 92.27\% is reached with a certainty of 95\%, in this scenario. This is also depicted by Figure \ref{fig:wSDhat_dist} B.

\subsection*{Retrospective assessment of test-retest studies}
It is not always necessary to conduct a preceding test-retest study when planning a longitudinal study. The $RC$ used in the longitudinal study might be adopted from already published test-retest studies. If one intends to use the point estimate of the $RC$ obtained in a previous study, it is advisable to  retrospectively assess the resulting distribution of the effective specificity and sensitivity. 
This allows to evaluate the impact of the sample size of the used test-retest study on quality criteria of the longitudinal study, especially the probability of exceeding a given $p_{\text{esp,lb}}$.\\
Common sample sizes in test-retest studies are around 10 and 20 \cite{barrett2019repeatability, giannotti2015assessment, miquel2012vitro, jerome2021understanding, barwick2021repeatability, michoux2021repeatability}. If the point estimator of $RC(0.95)$ resulting from a test-retest study with a sample size of 10 and two repeated measurements is used, the distribution of the effective specificity will have prominent tails as illustrated in Figure \ref{fig:bias_and_tails}. According to  \eqref{eq:esp_lb_exact}, the lower bound of the effective specificity obtained with 95\% confidence is 0.7814 and 0.8512 for a sample size of 10 and 20, respectively,  which might be insufficient (Figure \ref{fig:Formel13}). Note that for the recommendation by Obuchowski and Bullen\cite{Obuchowski2018-bo} of a sample size of 35 for test-retest studies with $m=2$ the probability of achieving an effective specificity below 94\% is 39.74\%.\\
Such considerations are also possible for the effective sensitivity.

%% file: discussion.tex
We have established a comprehensive framework for planning of test-retest studies concerning repeatability. It enables flexible calculation of sample size requirements and retrospective assessment of such studies with regard to different quality criteria.\\
To better discuss planning of test-retest studies we have introduced the notions of effective specificity ($P_{\text{esp}}$) and effective sensitivity ($P_{\text{ese}}$), allowing for clearer differentiation of the targeted specificity $p_{\text{sp}}$ and sensitivity $p_{\text{se}}$ from the values actually achieved in the longitudinal study. Both $P_{\text{esp}}$ and $P_{\text{ese}}$ are random quantities and their actual values are unknown in practical application. However, we can determine their distribution and thus can compute different characteristics which properly reflect the uncertainty caused by the estimation process. \\
Expanding on the work of Obuchowski and Bullen \cite{Obuchowski2018-bo}, we have introduced a new quality criterion for sample size calculation of test-retest studies. In their work, Obuchowski and Bullen\cite{Obuchowski2018-bo} demand that the mean effective specificity ($\mathbb{E}[P_{\text{esp}}]$) deviates at most by 0.01 from the fixed targeted specificity ($p_{\text{sp}}$) of 0.95.
However, using the mean effective specificity as sole quality criterion has limitations, since the whole distribution of the effective specificity is not properly taken into account. As illustrated in Figure \ref{fig:bias_and_tails}, there is a high probability that the actually achieved effective specificity deviates strongly from its target even if the mean effective specificity may be close to the targeted specificity. Therefore, we propose a quality criterion for sample size calculations based on the probability that the effective specificity exceeds a chosen lower bound, taking into account the tails of the distribution of $P_{\text{esp}}$.\\
In contrast to previous works we expand our consideration also to issues of sensitivity. Here, of course, it must also be taken into account that the sensitivity depends on the underlying effect size. Nevertheless, we can determine the distribution of the effective sensitivity for any effect size and provide analogous sample size formulas as for the specificity.
\\
Finally, our study is the first to provide analytical rather than simulation results. This provides greater flexibility as the targeted specificity $p_{\text{sp}}$ and number of repeated measurements $m$ may be chosen freely. Hence, it allows the readers to avoid conducting time-consuming simulation studies themselves.  
While our formulas enable flexible calculations for all scenarios, for convenience of the reader we also provide a table with sample sizes for some exemplary scenarios in Figure \ref{fig:Formel13}. Sample sizes resulting from other choices of the parameters $p_{\text{conf}}$, $p_{\text{esp,lb}}$, $p_{\text{sp}}$ and $m$ can be found in Supplementary Tables S1-S4.\\   
Our study has some limitations. The field of application is restricted to test-retest studies in which true replicates of measurements are possible,  for example in quantitative imaging markers. Our considerations are not valid if the measurement process itself results in a change of the measurand (learning / practice effect) as has been described for some psychological assessments \cite{hinton2011specificity, maassen2010two}.\\
Our standard model \eqref{eq:standard_model} assumes independent and identically normally distributed errors. 
It is therefore advisable to examine whether there is a relationship between the within-subject variation and the level of the measured value before applying our approach \cite{obuchowski2018interpreting}.  If the variability of the measurement error increases with the magnitude of the measured value, a log transformation might resolve the issue \cite{altman1983measurement, bland1996statistics, obuchowski2018interpreting}.
Beyond that, non-normally distributed error terms are not covered so far. We also have not specifically considered the scenario of clustered data, e.g. measuring multiple lesions per subject. However, if a hierarchical model structure with independent errors can be assumed, this does not pose a restriction to application of our approach.\\
It should be noted that exact solutions based on the $\chi^2$ distribution for all our considerations are available. In some cases, when an analytic solution is not possible, these exact solutions require the application of numerical methods. In order to give completely analytic solutions, some of our formulas rely on asymptotic results and approximations. The differences between exact and approximate results are most severe for small sample sizes and small effect sizes. Applying both exact and approximate formulas in our application example, 
it can be seen that these differences are negligible in practically relevant scenarios. Implementations of exact and approximate solutions can be found in our supplementary R code \cite{R}.\\
So far, our considerations are limited to repeatability, i.e. assuming same measurement conditions for the repeated measurements. However, for real world application of biomarkers, consideration of reproducibility is also important since longitudinal measurements are often performed under different measuring conditions, e.g. varying readers or scanners. Therefore, our model should be perspectively enhanced to include aspects of reproducibility such as a fixed bias as e.g. in some models considered by Obuchowski and Bullen \cite{Obuchowski2018-bo}. Nevertheless, since repeatability limits reproducibility, a good knowledge of the former is useful in order to interpret reproducibility studies properly \cite{bland1999measuring}.\\  
Test-retest studies of repeatability should be well planned to guarantee for a sufficient quality of dependent longitudinal studies. Our framework allows the derivation of analytical solutions for quality criteria that can be used to assess implications of the test-retest study design on subsequent longitudinal studies.